\documentstyle[eqsecnum,prd,aps,floats,epsf]{revtex}
\def\ee{e^{+}e^{-}}

\begin{document}
\onecolumn
\title{Search for Quark-Lepton Compositeness at Tevatron and LHC} 
\author{
Supriya Jain \and Ambreesh K. Gupta \and Naba K. Mondal}

\address{ Department of High Energy Physics, \\
Tata Institute of Fundamental Research, Homi Bhabha Road, \\Mumbai 400 005, 
India.}

\date{\today}
\maketitle

\begin{abstract}
We make a Monte Carlo study on compositeness of first generation quarks and 
leptons using the Drell-Yan distribution in the high dielectron mass region 
at the Tevatron and LHC energies. The current experimental lower limits on the 
compositeness scale, $\Lambda$, vary from 2.5 to 6.1 TeV. In the present 
analysis, we assume that there will be no deviation of the dielectron mass 
spectrum from Standard Model prediction at center of mass energy 2 TeV 
(Tevatron) and 14 TeV (LHC). We then find that in the LL, RR, RL and LR 
chirality channels of the quark-electron currents, it is possible to extend 
the lower limits on $\Lambda$ (at 95$\%$ {\it CL}) to a range of 6 to 10 TeV 
for $2~fb^{-1}$ and 9 to 19 TeV for $30~fb^{-1}$ of integrated luminosity at 
Tevatron. At LHC, the corresponding limits extend to a range of 16 to 25 TeV 
for $10~fb^{-1}$ and 20 to 36 TeV for $100~fb^{-1}$ of integrated luminosity.\\

\end{abstract}


\twocolumn

The proliferation of quarks and leptons has inspired the speculation that 
they could be composite structures, i.e. bound states of more fundamental 
constituents often called preons \cite{lane}. Below a characteristic energy 
scale called the compositeness scale, $\Lambda$, the preon-binding interaction
becomes strong and binds the constituents to form composite states like the 
quarks and leptons. With such a composite structure, there would be significant
deviation from the Standard Model (SM) prediction of high energy cross 
sections. No such deviation has been observed so far. These null results have
been used to put lower limits on quark-lepton compositeness scale $\Lambda$. 
which varies from 2.5 to 6.1 TeV \cite{lim} in the various chirality channels 
of the quark-lepton currents.\\

In this paper, we consider the effects of composite structure of first 
generation quarks and leptons on the Drell-Yan (DY) process 
$q {\bar{q}}~{\rightarrow}~ e^+ e^- $ \cite{drell}. If the compositeness 
scale, $\Lambda$, is much greater than $\sqrt{\hat s}$, the center of mass 
energy of the colliding partons, the quarks and electrons 
would appear to be point-like. The substructure coupling can then be 
approximated by a four-fermion contact interaction giving rise to the 
following effective lagrangian{\footnote{\footnotesize Here we have assumed 
that the contact interaction is color singlet and weak-isoscalar.}} 
\cite{lane}:\\

{\footnotesize{
\begin{eqnarray}
\label{eq:star1}{\mathcal{L}}_{ql} & = &  \frac{g^2_0}{\Lambda^2} 
 \bigg\{ {\eta_{LL}} ({\bar{q}}_L \gamma^\mu q_L)({\bar{e}}_L
                     \gamma_\mu e_L) + {\eta_{LR}}({\bar{q}}_L \gamma^\mu q_L)(                     {\bar{e}}_R
                     \gamma_\mu e_R) \\\nonumber
                   &  & +~~ \eta_{RL}({\bar{u}}_R \gamma_\mu u_R)({\bar{e}}_L \gamma                       ^\mu e_L) + {\eta_{RL}}({\bar{d}}_R
                     \gamma_\mu d_R)({\bar{e}}_L \gamma^\mu e_L) \\\nonumber
                   &  & +~~ \eta_{RR}({\bar{u}}_R \gamma^\mu u_R)({\bar{e}}_R \gamma
                     _\mu e_R) + {\eta_{RR}}({\bar{d}}_R \gamma^\mu d_R)
                     ({\bar{e}}_R\gamma_\mu e_R) \bigg\}\nonumber
\end{eqnarray}
}}
where
\begin{eqnarray}
\nonumber q_{L}  = \left[ \begin{array}{c}
                     u \\
                     d
                    \end{array} \right]_{L}  
\end{eqnarray}
is the left-handed quark doublet. $u_R$ and $d_R$ are the right-handed quark 
singlets. $e_L$ and $e_R$ are the left- and right-handed electrons 
respectively. The compositeness scale $(\Lambda) $ is 
chosen so that $\frac{g^2_0}{4\pi}=1 $ and the largest 
$\mid{\eta}_{ij}\mid=1$, where $g_0$ is the coupling constant for the contact 
interaction and $\eta_{ij} $ is the interference term between the contact 
interaction and the SM lagrangian for the $ij^{th} $ 
channel, with $\mathit i$ and $\mathit j$ representing the helicities of 
the quark  and the lepton currents. Including the 
above contact interaction (at $\Lambda >> \sqrt{\hat s}$), the DY cross 
section gets transformed as \cite{taekoon}:
\begin{eqnarray}
\label{eq:star9}\frac{d\sigma^\Lambda}{dm}~ =~ \frac{d\sigma}{dm}(DY) + {\beta}I + 
{{\beta}^2}C,
\end{eqnarray}
where $\beta = 1/\Lambda^2 $ and m is the dielectron invariant mass. 
In this expression, $\mathit I$ is due to the interference of DY and 
the contact 
term, and $\mathit C$ is the pure contact term contribution to the 
cross-section. The deviation in the dielectron production from SM 
expectations would be dominant in the high mass region above the Z pole. We 
have made separate studies for quark-
electron compositeness for an integrated luminosity of 2 $fb^{-1}$ (Run II) 
and 30 $fb^{-1}$ (TEV33) with respect to the D$\O$ detector at Tevatron and 
an integrated luminosity of 10 $fb^{-1}$ and 100 $fb^{-1}$ with respect to 
the CMS detector at LHC. However the results should be valid for the CDF 
detector at Tevatron and the ATLAS detector at LHC as well. We have 
simulated dielectron production through DY process alone in {\em p$\bar p$} 
({\em pp}) collisions at center of mass energy, $\sqrt{s}$, equal to 
2 TeV (14 TeV) using PYTHIA \cite{pythia}. However since PYTHIA does 
not incorporate all the compositeness models, we have used a separate 
parton level Monte Carlo program to estimate dielectron production rates 
in the presence of compositeness. Assuming that the Tevatron and LHC data on 
dielectron production are consistent with DY predictions under SM, we 
extract limits on compositeness scale using Bayesian technique of 
statistical inference \cite{baye1,d0baye}.  We have considered four 
different models corresponding to the LL, RR, RL and LR 
chirality channels of equation \ref{eq:star1} for quark-electron 
compositeness. The choice of $\eta_{ij} $ for the different models of 
compositeness is listed in Table~\ref{tab:table1}.\\
 
\begin{table}[h]
\caption{Choice of $\eta_{ij}$ for different contact 
interaction models. The superscript on the model denotes the nature of 
interference between the contact interaction and the SM lagrangian. 
Constructive interference $(\eta_{ij}~=~-1)$ is denoted by a $+$ and 
destructive interference $(\eta_{ij}~=~+1)$ is denoted by a $-$.} 
\label{tab:table1}
\begin{center}
\begin{tabular}{c c c c c}
Model & $\eta_{LL}$ & $\eta_{RR}$ & $\eta_{LR}$ & $\eta_{RL}$ \\\hline
$LL^{\pm}$ & $\mp 1$ & 0 & 0 & 0 \\
$RR^{\pm}$ & 0 & $\mp 1$ & 0 & 0 \\ 
$LR^{\pm}$ & 0 & 0 & $\mp 1$ & 0 \\
$RL^{\pm}$ & 0 & 0  & 0 & $\mp 1$ \\ 
\end{tabular}
\end{center}
\end{table}

\hspace*{-3mm}{\it Exploring the lower limits on $\Lambda$ at Tevatron}\\

We simulate {\em p$\bar p$} collisions using PYTHIA at 2 TeV and generate DY 
dielectron events between 95 GeV and 1.5 TeV of the dielectron invariant mass. The total number of dielectron events generated by PYTHIA, $N_{gen}$, gives 
the expected number of background subtracted dielectron events, $N_{DY}$, to 
be collected at Tevatron as :

\begin{eqnarray}
N_{DY} = \epsilon \times N_{gen}\label{eq:star5}
\end{eqnarray}
where {\em $\epsilon$} is the detection efficiency of the dielectron. The 
detection efficiency, $\epsilon$, of the dielectron involves 
contribution from the following terms: 
\begin{itemize}
\item[(a)] Energy smearing,
\item[(b)] Electron identification efficiency, $\epsilon_1$, and
\item[(c)] Acceptance, $\epsilon_2$.
\end{itemize}
The energy resolution of the electromagnetic calorimeter of the upgraded 
D$\O$ detector is parameterized as :
\begin{eqnarray}
(\frac{\sigma}{E})^2 =  C^2 + (\frac{a}{\sqrt{E}})^2   ~~~ (E~ in~ GeV)
\end{eqnarray}
where the constant term, C, and the stochastic term, $a$ are taken to 
2 $\%$ and 16 $\%$ respectively . We take the electron identification 
efficiency, $\epsilon_1$, for a single 
electron to be 85$\%$. The identification efficiency for a 
dielectron is then $\epsilon_1^2$. The acceptance, $\epsilon_2$, of 
dielectron events in {\em p$\bar p$} collisions 
is defined as the fraction of events in which the $\ee$ pair passes the 
fiducial and the kinematic cuts after taking into account the energy 
smearing. The fiducial and the kinematic cuts used are:
\begin{itemize}
\item $\mid{\eta}\mid \leq 2.5$, where $\eta$ is the pseudorapidity 
($=-ln[tan(\frac {\theta} {2})]$). This ensures that the dielectron event 
selected is in the active detector region. 
\item A kinematic cut of $p_T \geq 25~ GeV$, where $p_T$ is the transverse 
momentum of the electron and the positron. This cut ensures an efficient 
trigger.\footnote{\footnotesize This cut is based on the D$\O$ Run I analysis 
of DY data at 1.8 TeV \cite{thesis}.}
\end{itemize}
The dielectron detection efficiency, $\epsilon$, is then :
\begin{eqnarray}
\epsilon ~= ~{\epsilon_1}^2~ {\times}~ {\epsilon_2}
\end{eqnarray}

We then generate the {\em expected} number of dielectron events, 
$N_{exp}^{\Lambda}$, in various mass bins including the effect of the 
composite structure of quarks and electrons for various values of $\Lambda$
using the parton level Monte Carlo. We calculate the 
cross section ($\sigma^{\Lambda}$) for the production of 
dielectrons including terms from the contact interaction lagrangian of 
equation \ref{eq:star1} with the SM lagrangian. The LO 
cross section calculation is corrected for higher order QCD effects using a 
K-factor of 1.22{\footnote{\footnotesize This K-factor is the ratio of the 
NNLO DY cross section to the LO DY cross section at 1.8 TeV 
\cite{neerven,neerven1}. We consider the same value for the K-factor for 
DY + compositeness at 2 TeV.}}. 
We checked the parton level MC calculation by comparing its
\begin{figure}[hbt]
\centerline{\epsfxsize=7.cm\epsfbox{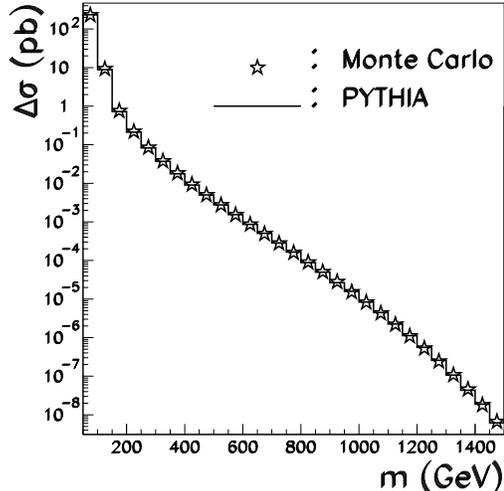}}
\caption{Dielectron invariant mass spectra between 80 GeV and 1.5 TeV for DY process at $\sqrt{s}~ =~ 2~TeV $, as predicted by PYTHIA and as calculated using our parton level Monte Carlo.}
\label{fig:reco6}
\end{figure}
prediction with that from PYTHIA for the Drell-Yan process. Both calculations 
agree to within a few percent as shown in Fig.~\ref{fig:reco6}.\\

\begin{figure}[hbt]
\centerline{\epsfxsize=7.0cm\epsfbox{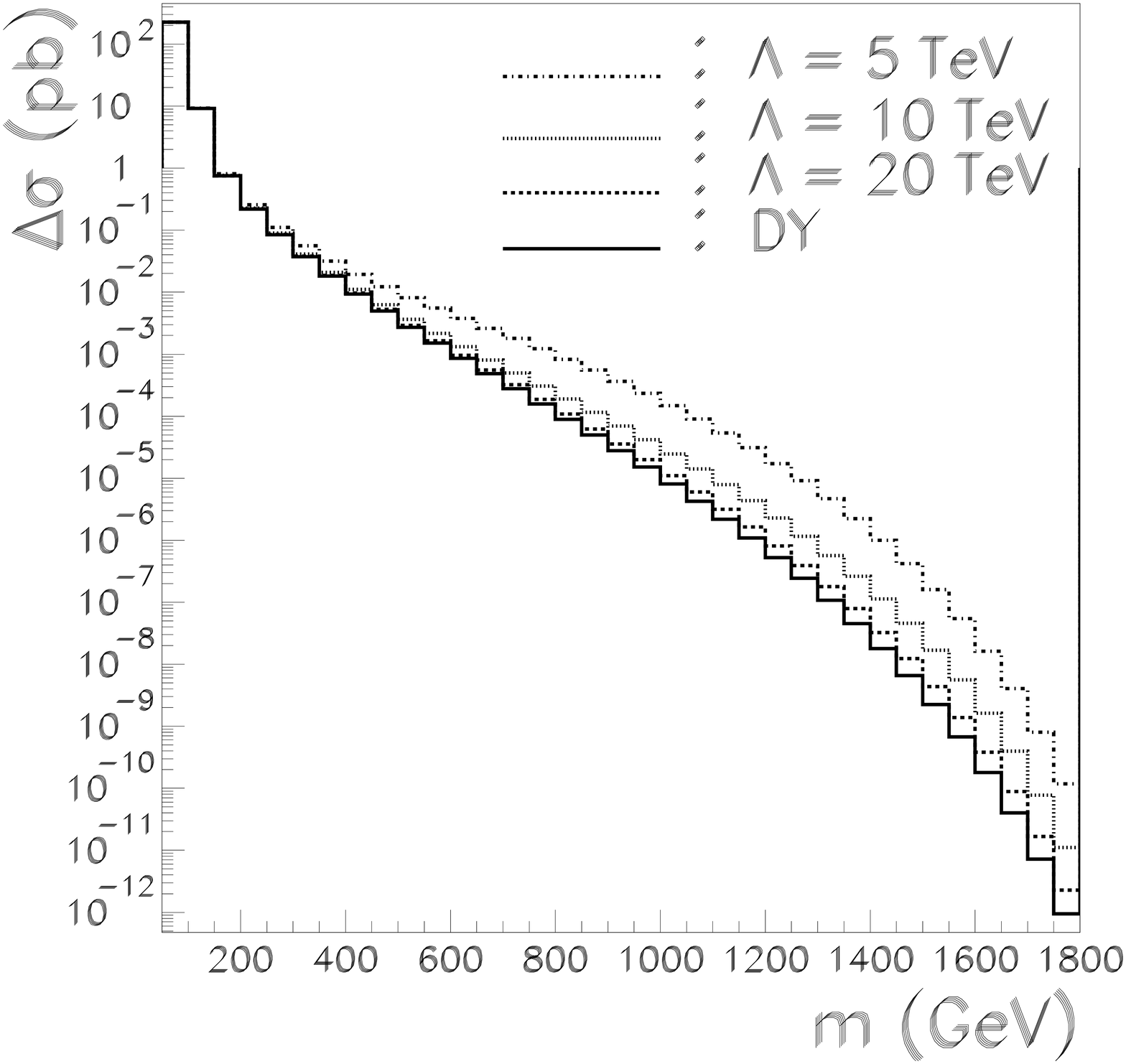}}
\caption{Cross section, $\Delta \sigma$ (in 50 GeV bins), versus dielectron invariant mass, m, between 50 GeV and 1.8 TeV for DY process and three different values of $\Lambda$ in the LL channel 
for $\eta_{ij}~=~-~1$.}
\label{fig:reco7}
\end{figure}

\begin{figure}[hbt]
\centerline{\epsfxsize=7.0cm\epsfbox{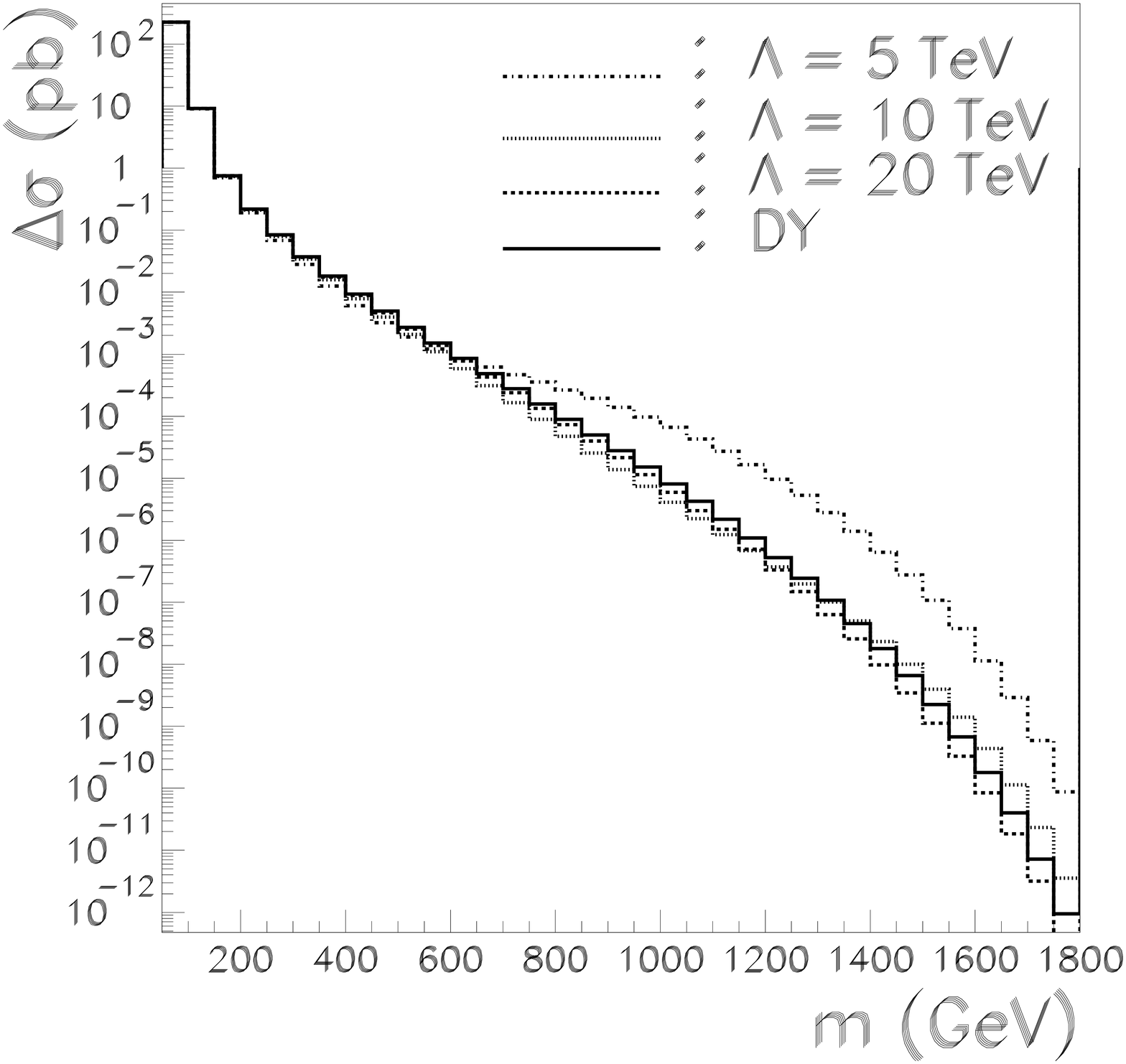}}
\caption{Cross section, $\Delta \sigma$ (in 50 GeV bins), versus dielectron invariant mass, m, between 50 GeV and 1.8 TeV for DY process and three different values of $\Lambda$ in the LL channel 
for $\eta_{ij}~=~+~1$.}
\label{fig:reco8}
\end{figure}

In order to obtain the lower limit on $\Lambda$, we then use the Bayesian 
technique to compare  the Drell Yan  dielectron mass distribution 
(ie., $N_{DY}$) in the high mass region with the expected dielectron mass 
distribution for various values of $\Lambda$ (ie., $N_{exp}^{\Lambda}$). 
Limits are obtained independently for each separate channel of the contact 
interaction lagrangian: LL, RR, RL and LR with $\eta_{ij} = \pm 1$. Fig.~\ref{fig:reco7} shows the cross section versus the dielectron invariant mass, in 
the high mass region between 50 GeV and 1.8 TeV in the LL channel for different values of $\Lambda$ for $\eta_{ij}~=~-~1$ (constructive interference) and 
Fig.~\ref{fig:reco8} shows the corresponding plot for 
$\eta_{ij}~=~+~1$ (destructive interference). \\

Since the effect of compositeness is most pronounced in the high dielectron 
mass region we consider 10 different mass bins of variable width between 120 
GeV and 1.5 TeV. The expected number of events at the compositeness scale, 
$\Lambda$, in the $k^{th}$ mass bin is given as :  
\begin{eqnarray}
\label{eq:repeat3}N_{exp}^{\Lambda,k} = \epsilon^k~ ( \sigma^{\Lambda,k} 
\times L ),
\end{eqnarray}
where $\sigma^{\Lambda,k}$ is the cross section (including 
compositeness) for the $k^{th}$ mass bin and $L$ is the integrated luminosity. 
The posterior probability for the compositeness 
scale to be $\Lambda$ given the expected DY dielectron data distribution, 
$d_O$, is:
\begin{eqnarray}
\label{eq:star7}P(\mathit \Lambda \mid \mathit d_O) = 
\frac{1}{\mathcal{Z}}\prod_{k~=~1}^n P^k(N_{DY}^k \mid N_
{exp}^{\Lambda,k} )~P(\epsilon^k,~L,~\Lambda)
\end{eqnarray}
where ${\mathcal{Z}}$ is the normalization constant. 
$P^k(\mathit N_{DY}^k \mid \mathit N_{exp}^{\Lambda,k})$ is the 
{\em likelihood} function which follows a 
Poisson distribution for small $N_{exp}^{\Lambda,k}$:
{\footnotesize
\begin{eqnarray}
P^k(N_{DY}^k \mid N_{exp}^{\Lambda,k}) = \frac {e^{-N_{exp}^{\Lambda,k}}~ 
(N_{exp}^{\Lambda,k})^{N_{DY}^k}}{N_{DY}^k !},~~~(N_{exp}^{\Lambda,k}~ <~10) 
\end{eqnarray}
}
\hspace*{-1.mm}and a Gaussian distribution for large $N_{exp}^{\Lambda,k}$, 
with mean $N_{exp}^{\Lambda,k}$ and standard deviation, $\sigma_1$, 
($\sigma_1~=~\sqrt{N_{exp}^{\Lambda,k}}~$) \cite{baye2} :
{\footnotesize
\begin{eqnarray}
P(N_{DY}^k \mid N_{exp}^{\Lambda,k}) = \frac {1}{\sqrt{2\pi}\sigma_1}~e^{-
{\frac{({N_{DY}^k-N_{exp}^{\Lambda,k}})^2}{2\sigma_1^2}}},~~~(N_{exp}^{\Lambda,k}~ \geq~10). 
\end{eqnarray} 
}
\hspace*{-1.mm}$P(\epsilon^k,~L,~\Lambda)$ is the joint $\mathit prior$ 
probability for the 
dielectron detection efficiency, $\epsilon^k$, the integrated luminosity, $\mathit{L}$, 
and the compositeness scale, $\Lambda$. Taking $\mathit \epsilon^k$, 
$\mathit L$ and $\Lambda$ to be independent,
\begin{eqnarray}
P(\epsilon^k, L, \Lambda) = P(\epsilon^k)~P(L)~P(\Lambda). 
\end{eqnarray}
The $\mathit {prior}$ probabilities of detection efficiency, 
$\mathit\epsilon^k$, and integrated luminosity, $\mathit L$, are assumed to be 
Gaussian 
with their estimated value in each bin as the $\mathit mean$ and 
corresponding error as the $\mathit width$ of the Gaussian. The prior 
distribution $P(\Lambda)$ is chosen to be uniform in $1/\Lambda^2$. This 
represents a prior essentially flat in cross section. The resulting posterior 
density $P(\Lambda \mid \mathit d_O)$ peaks at $1/\Lambda^2~=~0$ and falls 
off monotonically with increasing $1/\Lambda^2$. The $95 \%$ CL lower limit 
on $\Lambda$ is defined by: 
\begin{eqnarray}
{\int_{\Lambda_{lim}}^{\infty}}d\Lambda'~ P(\Lambda' \mid \mathit d_O ) = 0.95.
\end{eqnarray}

The values of efficiency, $\epsilon^k$, and the expected number of DY events, 
$N_{DY}^k$\footnote{\footnotesize $N_{DY}^k$ is generated with a K-factor of 1.22 in PYTHIA.}, in 
individual mass bins are listed in Table~\ref{tab:table2} for an integrated 
luminosity of 2 $fb^{-1}$ and 30 $fb^{-1}$. The expected 95$\%$ CL lower 
limits on $\Lambda$ for the LL, RR, RL and LR helicity channels of the quark-
electron currents for both constructive and destructive interference are 
listed in Table~\ref{tab:table3} and Table~\ref{tab:table4} for integrated 
luminosities of 2 $fb^{-1}$ and 30 $fb^{-1}$ 
respectively. \\  
\begin{table}[h]
\caption{Detection efficiency and expected number of DY events in different 
mass bins} 
\label{tab:table2}
\begin{center}
\begin{tabular}{c c c c} 
mass bin & $\epsilon^k$ & $N_{DY}^k$ & $N_{DY}^k$ \\
$ (GeV)$ & $~~~~$ & $L~=~2~fb^{-1}$ & $L~=~30~fb^{-1}$ \\  \hline
120-160 & 0.590 & 2335.8 & 34508.1  \\
160-200 & 0.629 & 606.9 & 8990.1  \\
200-240 & 0.655 & 236.3 & 3589.4  \\
240-290 & 0.663 & 117.8 & 1942.8  \\
290-340 & 0.675 & 66.5 & 877.1 \\
340-400 & 0.668 & 34.0 & 461.7  \\
400-500 & 0.689 & 23.8 & 276.0  \\
500-600 & 0.712 & 6.5 & 98.3  \\
600-1000 & 0.677 & 1.5 & 42.6  \\
1000-1500 & 0.723 & 0 & 2.2  \\
\end{tabular}
\end{center}
\end{table}
\begin{table}[h]
\caption{Expected 95$\%$ CL lower limits, $\Lambda_{lim}$, on the compositeness scale for different helicity channels of the quark-electron 
currents for $L~ = ~2 ~fb^{-1}$ at 2 TeV with $ \delta\epsilon^k~=~15~\%$ and  $\delta L~=~5~\%$}
\label{tab:table3}
\begin{center}
\begin{tabular}{c c c} 
$~~~~$ & $\Lambda_{lim}~(TeV)$ & $\Lambda_{lim}~(TeV) $ \\
Channel & $(\eta_{ij}~=~-~1)$ & $(\eta_{ij} ~=~+~1)$ \\ \hline
LL & 10.1 & 8.0     \\
RR & 9.3 & 6.0    \\
RL & 7.8 & 5.7     \\
LR & 7.3 & 6.0    \\ 
\end{tabular}
\end{center}
\end{table}

\begin{table}[h]
\caption{Expected 95$\%$ CL lower limits, $\Lambda_{lim}$, on the compositeness scale for different helicity channels of the quark-electron 
currents for $L~ = ~30 ~fb^{-1}$ at 2 TeV with $ \delta\epsilon^k~=~15~\%$ and  $\delta L~=~5~\%$}
\label{tab:table4}
\begin{center}
\begin{tabular}{c c c}
$~~~~$ & $\Lambda_{lim}~(TeV)$ & $\Lambda_{lim}~(TeV) $ \\
Channel & $(\eta_{ij}~=~-~1)$ & $(\eta_{ij} ~=~+~1)$ \\ \hline
LL & 18.9 & 17.8      \\
RR & 17.0 & 15.1    \\
RL & 13.5 & 9.1    \\
LR & 12.1 & 9.2     \\ 
\end{tabular}
\end{center}
\end{table}

\hspace*{-3mm}{\it Exploring the lower limits on $\Lambda$ at LHC} \\

\begin{figure}[hbt]
\centerline{\epsfxsize=7.0cm\epsfbox{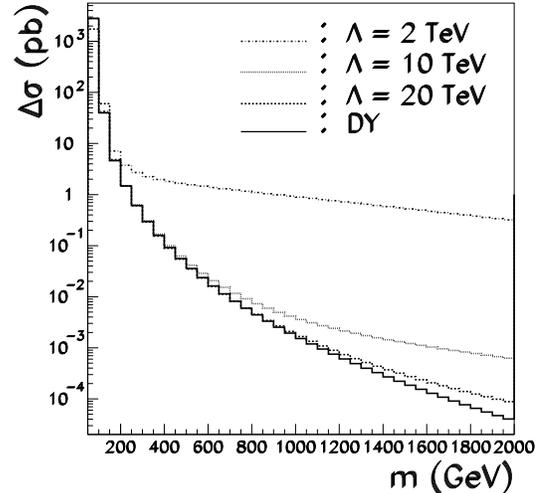}}
\caption{Cross section, 
$\Delta\sigma$ (in 50 GeV bins), versus dielectron invariant mass, m, between 
50 GeV and 2 TeV for DY process and three different values of $\Lambda$ in 
the LL channel for $\eta_{ij}~=~-~1$.}
\label{fig:lhclam+}
\end{figure}

\begin{figure}[hbt]
\centerline{\epsfxsize=7.0cm\epsfbox{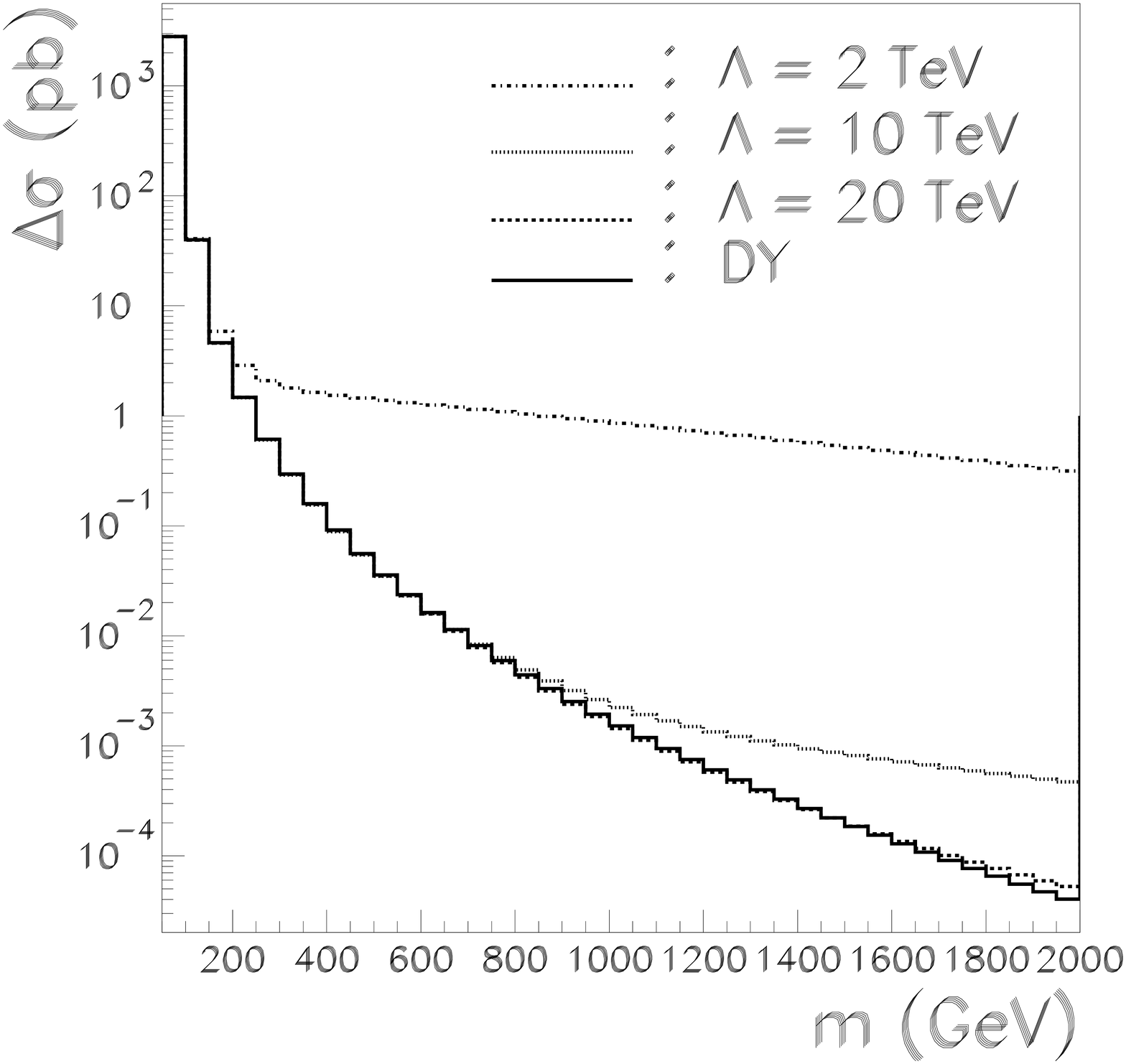}}
\caption{Cross section, 
$\Delta \sigma$ (in 50 GeV bins), versus dielectron invariant mass, m, 
between 50 GeV and 2 TeV for DY process and three different values of 
$\Lambda$ in the LL channel for $\eta_{ij}~=~+~1$.}
\label{fig:lhclam-}
\end{figure}

We have made a similar analysis of the DY process including the effect of 
quark-electron compositeness at 14 TeV. As before we have assumed that DY 
dielectron data that would be collected by the CMS detector at LHC would agree with SM prediction. We then use the Bayesian technique to obtain the lower 
limits on $\Lambda$ at 14 TeV. We have made separate studies for $10~fb^{-1}$ 
of data and $100~fb^{-1}$ of data. A K-factor of 1.13 \cite{neerven,neerven1} has been 
used as the NNLO correction factor. Fig.~\ref{fig:lhclam+} shows the cross 
section versus the dielectron invariant mass, in the high mass region between 
50 GeV and 2 TeV in the LL channel for different values of $\Lambda$ for 
$\eta_{ij}~=~-~1$ (constructive interference) and Fig.~\ref{fig:lhclam-} 
shows the corresponding plot for $\eta_{ij}~=~+~1$ (destructive interference). \\

We generated DY events in the dielectron mass range of 150 GeV to 2 TeV. We 
then compared the expected number of DY events, $N_{DY}$, at $\sqrt{s}$ = 14 
TeV with the expected number of dielectron events, $N_{exp}^{\Lambda}$, at 
various values of $\Lambda$ in the mass range of 500 GeV to 2 TeV where the 
deviation from SM predictions due to the composite structure of quarks and 
electrons is most pronounced at LHC. The electron identification efficiency, 
$\epsilon_1$, is taken to be 95$\%$ \cite{rep2}. The constant and stochastic 
terms in the energy resolution of the electromagnetic calorimeter of the CMS 
detector are taken to be \cite{rep2}:
\begin{eqnarray}
C & = & 0.55 \%,~~ and \\\nonumber
a & = & 2.7 \%,~ \mid \eta \mid~ \leq~ 1.5 \\\nonumber
& & 5.7 \%,~~ 1.5~~ ~<~ \mid \eta \mid~ \leq~ 2.5 
\end{eqnarray}

The fiducial and kinematic cuts selected are the same as for D$\O$. The values of $\epsilon^k$ and $N_{DY}^k$\footnote{\footnotesize $N_{DY}^k$ is generated 
with a K-factor of 1.13 in PYTHIA.}, in 
individual mass bins are listed 
in Table~\ref{tab:table5} for integrated luminosities of 10 $fb^{-1}$ and 100 
$fb^{-1}$.
\begin{table}[h]
\caption{Detection efficiency and expected number of DY events in different 
mass bins }
\label{tab:table5}
\begin{center}
\begin{tabular}{c c c c} 
mass bin & $\epsilon^k$ & $N_{DY}^k$ & $N_{DY}^k$ \\
$ (GeV)$ & $~~~~$ & $L~=~10~fb^{-1}$ & $L~=~100~fb^{-1}$ \\  \hline
500-510 & 0.660 & 56.86 & 586.63  \\
510-520 & 0.617 & 44.22 & 506.30  \\
520-530 & 0.668 & 48.74 & 493.67  \\
530-540 & 0.654 & 44.22 & 425.98  \\
540-550 & 0.647 & 37.91 & 418.76  \\
550-560 & 0.662 & 46.03 & 378.15  \\
560-570 & 0.666 & 37.91 & 361.90  \\
570-580 & 0.668 & 34.30 & 329.41  \\
580-600 & 0.673 & 45.13 & 552.33  \\
600-625 & 0.684 & 57.76 & 621.82  \\
625-650 & 0.678 & 50.54 & 509.91  \\
650-675 & 0.681 & 45.13 & 419.66  \\
675-700 & 0.692 & 28.88 & 386.27  \\
700-750 & 0.717 & 52.35 & 589.33  \\
750-800 & 0.728 & 36.10 & 452.15  \\
800-900 & 0.731 & 53.25 & 613.70  \\
900-1000 & 0.756 & 31.59 & 336.63  \\
1000-1200 & 0.752 & 36.10 & 315.88  \\
1200-1400 & 0.782 & 16.25 & 151.62  \\
1400-2000 & 0.791 & 16.25 & 135.38  \\  
\end{tabular}
\end{center}
\end{table}
The expected 95$\%$ CL lower limits on $\Lambda$ for the LL, RR, RL and LR 
helicity channels of quark-electron currents for both constructive and 
destructive interference are listed in Table~\ref{tab:table6} and Table~\ref{tab:table7} for integrated luminosities of 
10 $fb^{-1}$ and 100 $fb^{-1}$ respectively. \\
\begin{table}[h]
\caption{Expected 95$\%$ CL lower limits, $\Lambda_{lim}$, on the compositeness scale for different helicity channels of the quark-electron 
currents for $L~ = ~10 ~fb^{-1}$ at 14 TeV with $\delta\epsilon^k~=~15~\%$ and  $\delta L~=~5~\%$}
\label{tab:table6}
\begin{center}
\begin{tabular}{c c c} 
$~~~~$ & $\Lambda_{lim}~(TeV)$ & $\Lambda_{lim}~(TeV) $ \\
Channel & $(\eta_{ij}~=~-~1)$ & $(\eta_{ij} ~=~+~1)$ \\ \hline
LL & 24.0 & 16.4     \\
RR & 24.0 & 16.5    \\
RL & 21.4 & 17.6     \\
LR & 21.7 & 17.4    \\ 
\end{tabular}
\end{center}
\end{table}
\begin{table}[h]
\caption{Expected 95$\%$ CL lower limits, $\Lambda_{lim}$, on the compositeness scale for different helicity channels of the quark-electron 
currents for $L~ = ~100 ~fb^{-1}$ at 14 TeV with $\delta\epsilon^k~=~15~\%$ and  $\delta L~=~5~\%$}
\label{tab:table7}
\begin{center}
\begin{tabular}{c c c}
$~~~~$ & $\Lambda_{lim}~(TeV)$ & $\Lambda_{lim}~(TeV) $ \\
Channel & $(\eta_{ij}~=~-~1)$ & $(\eta_{ij} ~=~+~1)$ \\ \hline
LL & 33.8 & 20.1      \\
RR & 33.7 & 20.2    \\
RL & 29.2 & 22.1    \\
LR & 29.7 & 21.8     \\ 
\end{tabular}
\end{center}
\end{table}
The discovery limits for $\Lambda$ (defined as a deviation of 5$\sigma$ from 
SM prediction) for the various models have been listed for integrated 
luminosities of $10 ~fb^{-1}$, $50 ~fb^{-1}$, $100 ~fb^{-1}$, $200 ~fb^{-1}$ 
and $500 ~fb^{-1}$ in Table~\ref{tab:table8} for $\eta_{ij} ~=~-~1$ and in 
Table~\ref{tab:table9} for 
$\eta_{ij} ~=~+~1$. 
\begin{table}[h]
\caption{$\Lambda_{5\sigma}$ for five different integrated 
luminosities for {\bf $\eta_{ij}~=~-~1$} at $\sqrt{s}~=$ 14 TeV with $\delta\epsilon^k~=~15~\%$ and  $\delta L~=~5~\%$}
\label{tab:table8}
\begin{center}
\begin{tabular}{c|c c c c c} 
\multicolumn{1}{c}{} &
\multicolumn{5}{c}{$\Lambda_{5\sigma}~(TeV)$}\\\cline{2-6}
Channel&$10~fb^{-1}$&$50~fb^{-1}$&$100~fb^{-1}$&$200~fb^{-1}$&$500~fb^{-1}$\\ \hline
LL&16.0&20.6&23.4&26.2&31.0\\
RR&16.0&20.5&23.3&26.2&30.8\\
RL&15.1&18.6&20.9&23.2&26.8\\
LR&15.1&19.1&21.1&23.5&27.1\\ 
\end{tabular}
\end{center}
\end{table}
\begin{table}[h]
\caption{$\Lambda_{5\sigma}$ for five different integrated 
luminosities for {\bf $\eta_{ij}~=~+~1$} at $\sqrt{s}~=$ 14 TeV with $\delta\epsilon^k~=~15~\%$ and  $\delta L~=~5~\%$} 
\label{tab:table9}
\begin{center}
\begin{tabular}{c|c c c c c} 
\multicolumn{1}{c}{} &
\multicolumn{5}{c}{$\Lambda_{5\sigma}~(TeV)$}\\\cline{2-6}
Channel&$10~fb^{-1}$&$50~fb^{-1}$&$100~fb^{-1}$&$200~fb^{-1}$&$500~fb^{-1}$\\ \hline
LL&12.4&14.9&15.9&17.1&18.3\\
RR&12.4&14.9&16.0&17.1&18.4\\
RL&13.1&15.8&17.3&18.5&20.2\\
LR&13.0&15.7&17.2&18.3&20.0\\
\end{tabular}
\end{center}
\end{table}

\begin{figure}[hbt]
\centerline{\epsfxsize=7.5cm\epsfbox{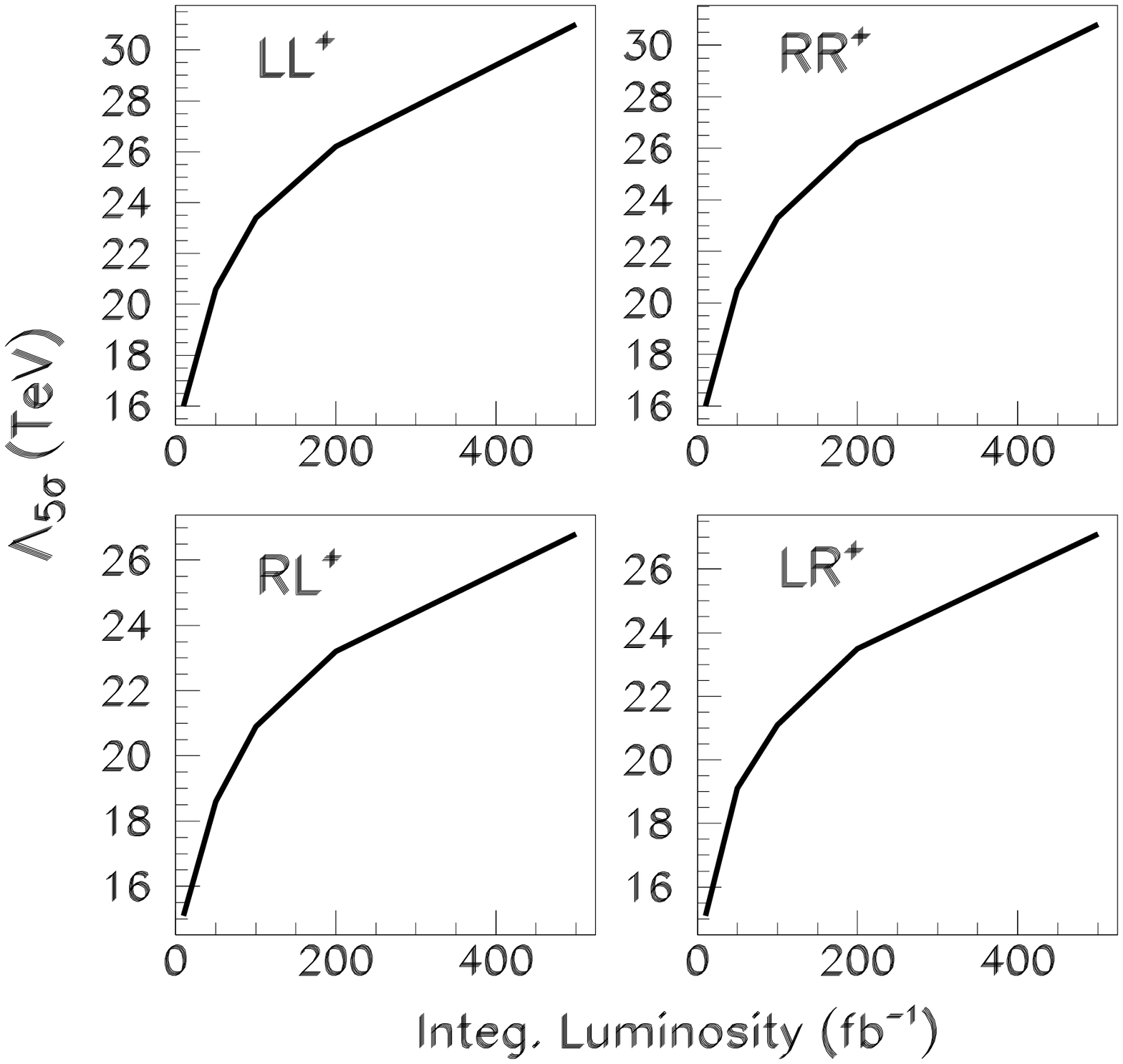}}
\caption{${5\sigma}$ discovery limit versus the integrated luminosity for $\eta_{ij}~=~-~1$ (constructive 
interference).}
\label{fig:reco13}
\end{figure}

\begin{figure}[hbt]
\centerline{\epsfxsize=7.5cm\epsfbox{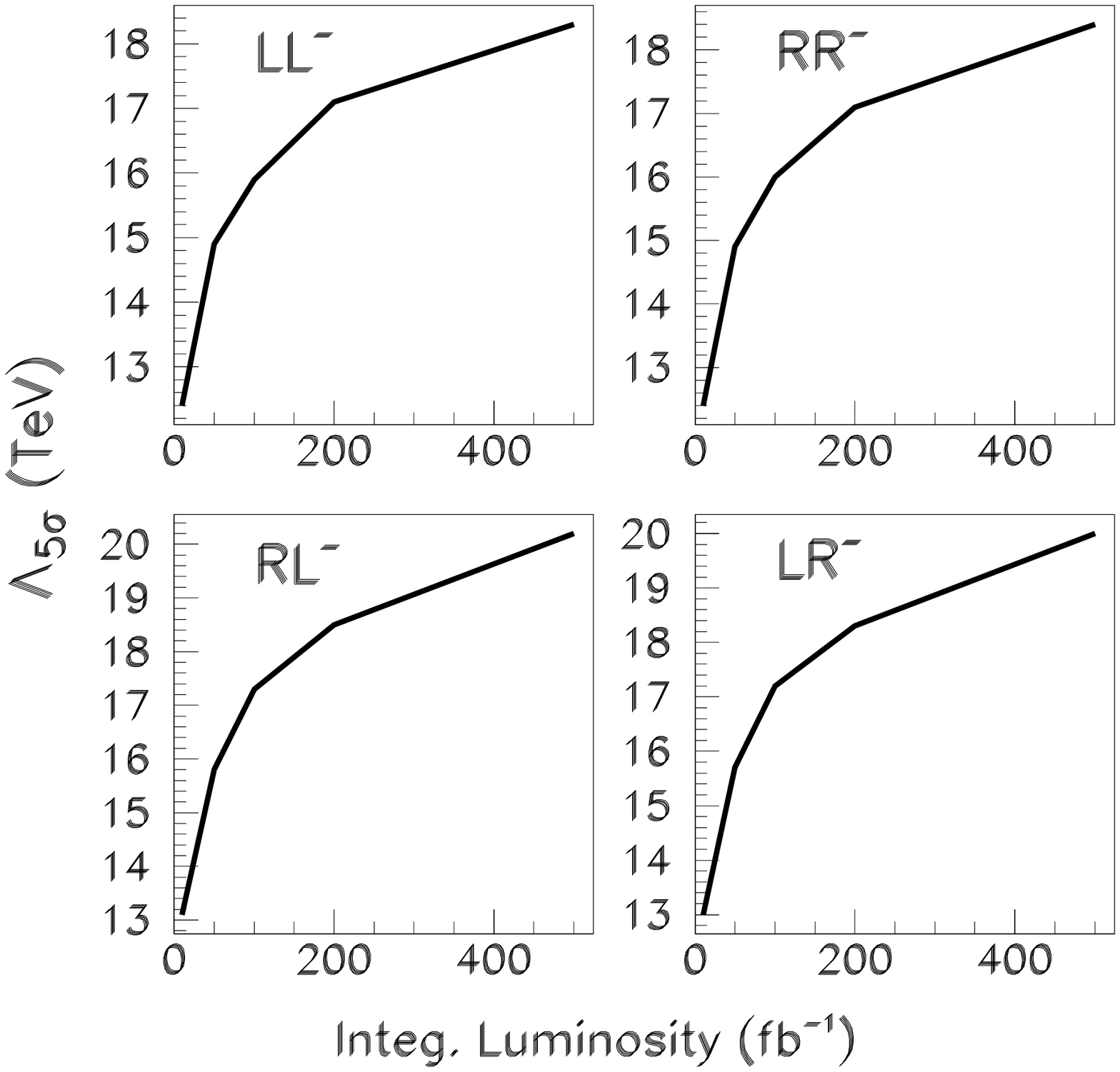}}
\caption{${5\sigma}$ discovery limit versus the integrated luminosity for $\eta_{ij}~=~+~1$ (destructive interference).}
\label{fig:reco14}
\end{figure}

Plots of the discovery limit versus the integrated luminosity for the various 
chirality channels are shown in Fig.~\ref{fig:reco13} for 
$\eta_{ij}~=~-~1$ and in Fig.~\ref{fig:reco14} for $\eta_{ij}~=~+~1$. \\

To conclude, we have performed a Monte Carlo study of the dielectron 
invariant mass spectrum (DY + compositeness) for $p \bar p$ collisions at 2 
TeV and $pp$ collisions at 14 TeV. We have considered the LL, RR, RL and LR 
chirality channels of the quark-electron currents. Assuming that 
Standard Model will describe the high mass DY dielectron data at 2 TeV and 
14 TeV we have found that it is possible to extend the lower limits on the 
compositeness scale, $\Lambda$, from the existing limits.\\
\begin{itemize}
\item For $p \bar p$ collisions at Tevatron we have made separate studies for 
integrated luminosities of $2~ fb^{-1}$ and $30 ~fb^{-1}$ with respect to the 
D$\O$ detector. The expected 95 $\%$ CL lower limits on $\Lambda$ range 
between 6 to 10 TeV and 9 to 19 TeV for $2~ fb^{-1}$ and $30 ~fb^{-1}$ of 
dielectron data, respectively. These limits are in agreement with similar 
limits on $\Lambda$ quoted between 6 to 10 TeV for $2~ fb^{-1}$ and 14 to 20 
TeV for $30 ~fb^{-1}$ of data with respect to the CDF detector at Tevatron 
\cite{pawel}.\\
\item For $pp$ collisions at LHC we have considered $10~ fb^{-1}$ and $100 ~fb^{-1}$ of dielectron data with respect to the CMS detector. The expected 95 $\%$ CL lower limits on $\Lambda$ range between 16 to 25 TeV for $10~ fb^{-1}$ and 
between 20 to 36 TeV for $100~ fb^{-1}$ of dielectron data. \\
\item We have also explored the discovery potential for quark-electron 
compositeness (defined as a deviation of 5$\sigma$ from SM prediction) at LHC 
as a function of integrated luminosity.
\end{itemize}

The authors would like to thank Sreerup Raychaudhury and V.S.Narasimham for 
their advice, comments 
and stimulating questions. We would also like to thank D.P.Roy and Sudeshna Banerjee for 
several fruitful discussions. 



\footnotesize
\begin{thebibliography}{1}
\bibitem{lane}E. Eichten,  K. Lane and M.Peskin, Phys. Rev. Lett. {\bf 50}, 
811 (1983);\\
E. Eichten, I. Hinchliffe, K. Lane and C. Quigg, {\em Review of 
Modern Physics},
 {\bf 56}, October 1984. 

\bibitem{lim}OPAL Collaboration, K. Ackerstaff {\em et al.}, Phys. Lett. B 
{\bf 391}, 221 (1997); CDF Collaboration, F. Abe {\em et al.}, Phys. Rev. Lett. {\bf 79}, 2198 (1997); D$\O$ Collaboration, B. Abbott {\em et al.}, Phys. Rev. Lett. {\bf 82}, 4769 (1999).

\bibitem{drell}S.D.Drell and T.M.Yan, Phys. Rev. Lett. {\bf 25}, 316 (1970).

\bibitem{taekoon}T. Lee, Phys. Rev. D {\bf 55}, 2591 (1997).

\bibitem{pythia}T. Sj$\o$strand, Comput. Phys. Commun. {\bf 82}, 74 (1994). 

\bibitem{baye1}G. Larry Bretthorst, Washington University, {\em An 
Introduction to Model Selection Using Probability Theory as Logic.}

\bibitem{d0baye}I.Bertram, G.Landsberg, J.Linneman, R.Partridge, M.Paterno, H.B.Prosper, {\em A Recipe for the Construction of Confidence Limits}, D$\O$ Note 2775A, December 18, 1995.  

\bibitem{thesis}Ambreesh Gupta, {\em Search for New Physics in} 
${\bar p}p~$ {\em Collisions at 1.8 TeV}, Ph.D thesis, Tata Institute 
of Fundamental Research, India, 1999.

\bibitem{neerven}R.Hamberg, W.L. van Neerven and T.Matsuura, Nucl. Phys. B 
{\bf 359}, 343 (1991).

\bibitem{neerven1}W.L. van Neerven, {\em Drell-Yan Production at Large Hadron 
Colliders}, INLO - PUB - 3/94. 

\bibitem{baye2} Giulio D ' Agostini, {\em Probability and Measurement 
uncertainty in Physics} - a Bayesian Primer (Notes based on lectures given to 
graduate students in Rome, May, 1995, and summer students at DESY, September, 
1995;

\bibitem{rep2}{\em The Electromagnetic Calorimeter Project}, 
Technical Design Report, CERN/LHCC 97-33,15 December 1997.

\bibitem{pawel}P.de Barbaro (CDF), {\em et al.}, Sensitivity to Compositeness Scale for $2~fb^{-1}$ and $30~fb^{-1}$ using Drell-Yan {\em ee} and $\mu\mu$ events at the Tevatron, Proceedings of Snowmass 96 Conference.

\end{thebibliography}
\end{document}